\documentclass[preprint,showpacs,preprintnumbers,amsmath,amssymb]{revtex4}
\usepackage{color}
\usepackage{graphicx}
\begin{document}

\preprint{KIAS-P13004}

\title{Higgs-dilaton(radion) system confronting the LHC Higgs data}



\author{Dong-Won Jung}
\email[]{dwjung@kias.re.kr}
\affiliation{School of Physics, KIAS, Seoul 130-722, Republic of Korea}

\author{P. Ko}
\email[]{pko@kias.re.kr}
\affiliation{School of Physics, KIAS, Seoul 130-722, Korea}


\date{\today}

\begin{abstract}
We consider the Higgs-dilaton(radion) system using the trace of the energy-momentum tensor 
($T_{~\mu}^\mu$) with the full Standard Model (SM) gauge symmetry 
$G_{\rm SM} \equiv SU(3)_c \times SU(2)_L \times U(1)_Y$, and find out that the 
resulting phenomenology for the Higgs-dilaton(radion) system is distinctly different from 
%
the earlier studies based on the $T_{~\mu}^\mu$ with the unbroken subgroup 
$H_{\rm SM} \equiv SU(3)_c \times U(1)_{\rm em}$ of $G_{\rm SM}$.  
After electroweak symmetry breaking (EWSB),  the SM Higgs boson and dilaton(radion) 
will mix with each other,  and there appear two Higgs-like scalar bosons and the  
Higgs-dilaton mixing changes the scalar phenomenology in interesting ways.
The signal strengths for the $gg$-initiated channels could be modified significantly 
compared with the SM predictions due to the QCD scale anomaly and the 
Higgs-dilaton(radion) mixing,  whereas anomaly contributions are almost negligible 
for other channels.  
We also discuss the self-couplings  and the signal strengths of the $126$ GeV scalar  
boson in various channels and possible constraints from the extra light/heavy scalar boson. 
The Higgs-dilaton(radion) system considered in this work has a number of distinctive 
features that could be tested by the upcoming LHC running and at the ILC. 
\end{abstract}


\maketitle

\section{Introduction}

Scale symmetry has been an interesting subject both in formal quantum field theory 
and in particle physics phenomenology~\cite{coleman:1988}. 
 The most important example is the scale 
invariance (Weyl invariance or conformal invariance) in string theory,  which is nothing 
but  2-dimensional quantum field theories for the string world sheet in target spacetime 
of spacetime dimensionality $d$.  The condition of vanishing quantum scale anomaly 
constrains possible perturbative string theories to be defined only in $d=26$ spacetime for 
bosonic string theory and $d=10$ spacetime for superstring theory. 
However, implementing scale symmetry to particle physics has not been so successful 
compared with string theory for various reasons. 

First of all, scale symmetry is always broken by  quantum radiative corrections through 
renormalization effects. Even if we start from a theory with classical scale symmetry 
(namely, no dimensional parameters in  ${\cal L}_{\rm classical}$),  
the corresponding quantum  theory always involves hidden scales, the cutoff scale 
($\Lambda$) in cutoff regularization or Pauli-Villars regularization,  and the renormalization
scale $\mu$ in dimensional regularization. 
In either case, scale symmetry is explicitly  broken by quantum effects, and 
scale symmetry is anomalous.  If the couplings do not run because of vanishing 
$\beta$ function, we would have truly scale invariant (or conformal symmetric) theory, and 
$N=4$ super Yang-Mills theory is believed to be such an example.

Secondly, scale symmetry may be spontaneously broken  by some nonzero values of 
dimensionful order parameters due to some nonperturbative dynamics, very often involving
some strong interaction. For example, we can consider massless QCD with classical scale
invariance. In this case there could be nonzero gluon condensate 
$\langle G_{\mu\nu}^a  G^{a\mu\nu} \rangle \sim \Lambda_{G^2}^4$  and chiral condensate 
$\langle \bar{q} q \rangle \sim \Lambda_{\bar{q}q}^3$, where new scales $\Lambda_{G^2}$ 
and $\Lambda_{\bar{q}q}$ are generated dynamically and they would be roughly of order of 
the confinement scale $\Lambda_{QCD}$.  
Since scale symmetry is spontaneously broken, there would appear  massless 
Nambu-Goldstone (NG) boson, which is often called dilaton related to dilatation 
symmetry.  If scale symmetry were not anomalously broken by quantum effects, dilaton 
could be exactly massless. However, scale symmetry is usually broken explicitly by 
renormalization effects, and dilaton would acquire nonzero mass which is related to the 
size of quantum anomaly, 
in a similar way to the pion as a pseudo Nambu-Goldstone boson in ordinary QCD.
If the dilaton mass is too large compared with the spontaneous scale symmetry breaking 
scale,  it is not meaningful to talk about dilaton as a pseudo NG boson. 
On the other hand, if dilaton is light enough, then we can use the nonlinear realization 
of scale symmetry with built-in quantum scale anomaly. 
Whether dilaton can be light enough or not is a very difficult question to address.  
The answer would depend on the underlying theories with classical scale symmetry, 
without which we cannot say for sure about pseudo NG boson nature of dilaton.  

Let us note that there have been longtime questions about generating the masses of 
(fundamental) particle only from quantum dynamics.  A good example is getting
proton mass from massless QCD. Since the contributions of  current quark masses 
to proton mass are negligible, we can say that proton mass is mostly coming from 
quantum dynamics between (almost massless) quarks and gluons.  Another well-known 
example is radiative symmetry breaking \`{a} la Coleman-Weinberg 
mechanism~\cite{Coleman:1973jx}.  In fact, 
a number of recent  papers address generating particle masses along this direction.
There are two different ways to getting mass scales from scale invariant classical theories:
one from new strong dynamics in a hidden sector~\cite{Hur:2007uz,Hur:2011sv,Holthausen:2013ota} 
and the other by CW mechanism 
~\cite{Foot:2007ay,Foot:2007iy,Meissner:2006zh,Iso:2009ss,Iso:2009nw,Holthausen:2009uc,Iso:2012jn,Chun:2013soa}.
If there are no mass parameters in classical Lagrangian, the theory would have classical 
scale symmetry. And all the mass scales would have been generated by quantum effects,
either nonperturbatively or perturbatively. 

Before the Higgs boson was discovered,  dilaton (denoted as $\phi$ in this paper) has been  considered as an alternative to the Higgs boson
~\cite{Goldberger:2007zk,Barger:2011hu,Yamawaki:1985zg,Appelquist:1986an,Foot:2007as}
from time to time, since dilation couplings to the SM fields are similar to the SM Higgs field at 
classical level,  except that the overall coupling scale is given by the dilaton decay constant 
$f_\phi$ instead of the Higgs vacuum expectation value (VEV) $v$.  
At quantum level, dilaton has couplings to the gauge kinetic functions due to the quantum 
scale anomaly~\cite{Coleman:1970je},  a distinct property of dilaton which is not shared 
by the SM Higgs boson. The radion~\cite{Goldberger:1999uk} in Randall-Sundrum (RS) 
model ~\cite{Randall:1999ee,Randall:1999vf} has the similar properties as the dilaton, 
in that it couples to the trace of the energy-momentum tensor too just like the dilaton 
\cite{Goldberger:1999uk,Giudice:2000av,Bae:2000pk,Csaki:2000zn}
\footnote{For this reason, we will use the terminology ``dilaton"  for both dilaton in 
spontaneously broken scale symmetric theory and the radion in the RS scenario 
in this paper.}. 


The interest in dilaton physics has been renewed recently 
\cite{Cheung:2011nv,Matsuzaki:2012gd,Matsuzaki:2012vc,Matsuzaki:2012mk,Matsuzaki:2012xx,Elander:2012fk,Hong:2013eta,
Chacko:2012sy,Chacko:2012vm,Bellazzini:2012vz,Coriano:2012nm,Abe:2012eu,Cao:2013cfa}, 
since the LHC announced discovery of a new boson of mass around 126~GeV 
(which we call  $H$ in this letter) 
\cite{ATLAS13-012,ATLAS13-013,ATLAS13-030,ATLAS13-079,ATLAS13-108,CMS13-001,Chatrchyan:2013mxa,Chatrchyan:2013iaa,Chatrchyan:2013zna,CMS13-004}. 
Radion-Higgs mixing scenarios have also been extensively studied in the light of the LHC results 
\cite{Toharia:2008tm,Grzadkowski:2012ng,deSandes:2011zs,Kubota:2012in,Desai:2013pga,Cox:2013rva}.
The current data still suffer from large uncertainties, but the observed new particle has 
properties that are consistent with the SM predictions, although there is a tendency that 
the $\gamma \gamma \left(Z Z^* \right)$ mode is enhanced over the SM predictions at 
ATLAS detector. The other modes are consistent with the SM predictions, but within a large uncertainty.

The effective interaction Lagrangian for a dilaton $\phi$ to the SM field can be derived 
by using nonlinear realization: $\chi={\rm e}^{\frac{\phi}{f_\phi}}$ \cite{coleman:1988}.
With the trace of the energy momentum tensor, which is the divergence of dilatation 
current, the  interaction terms which are linear in $\phi$ cast into 
\begin{widetext}
\begin{equation}
{\cal L}_{\rm int} \simeq -\frac{\phi}{f_\phi}~T_{~\mu}^\mu =- 
\frac{\phi}{f_\phi} \left[ m_h^2 h^2 - 2 m_W^2 W^+ W^- - 
m_Z^2 Z_\mu Z^\mu + \sum_{f} m_f \bar{f} f   + \sum_G
\frac{\beta_G}{g_G} G_{\mu\nu} G^{\mu\nu}   \right] ,
\end{equation}
\end{widetext}
where $m_h$ is the Higgs mass in the broken phase of the SM gauge group.
We argue that this form of dilaton interaction to the SM fields may not be proper, 
since only the unbroken subgroup of 
the SM gauge symmetry has been imposed on $T_{~\mu}^\mu$. If we imposed the 
full SM gauge symmetry on $T_{~\mu}^\mu$, the more proper form of the dilaton couplings 
to the SM should be described by Eq. (3) below, which is completely different from Eq. (1).

The SM Lagrangian is written as 
\begin{widetext}
\begin{equation}
{\cal L}_{\rm SM} = {\cal L}_{kin} ( G ) +  {\cal L}_{kin} ( f ) +  {\cal L}_{kin} ( H ) 
+ {\cal L}_{\rm Yukawa} ( f, \bar{f}, H )
- \mu_H^2 H^\dagger H - \lambda \left( H^\dagger H \right)^2,
\end{equation}
\end{widetext}
where $G$, $f$ and $H$ denote the SM gauge fields, fermions and Higgs 
field in a schematic way.
In this form, scale symmetry is explicitly broken by a single term, $\mu_H^2 H^\dagger H$
in the SM. Also quantum mechanical effects break scale symmetry anomalously.
In the end, the trace of energy-momentum tensor of the SM, which measures 
the amount of scale symmetry breaking, is given by 
\begin{equation}
T^\mu_{~\mu} ({\rm SM}) = 2\mu_H^2 H^\dagger H + \sum_G
\frac{\beta_G}{g_G} G_{\mu\nu} G^{\mu\nu} .
\end{equation}
This form of $T_{~\mu}^\mu$ respects the full SM gauge symmetry $G_{\rm SM} = 
SU(3)_C \times SU(2)_L \times U(1)_Y$. 
This form is clearly different from the usual form, Eq.~(1), which is constructed after 
EWSB and respects only the unbroken subgroup of the SM, $H_{\rm SM} = SU(3)_C 
\times U(1)_{\rm em}$. We claim that one has to use the form before EWSB, since we
do not know the scale of spontaneous scale symmetry breaking.  
If $v_{\rm EW} < f_\phi$, it would be more reasonable to impose the full SM gauge 
symmetry with Eq. (3)
\footnote{This form of the trace of energy-momentum tensor was considered in 
Refs.~\cite{Buchmuller:1987uc,Buchmuller:1990pz}. 
In these papers, the dilaton mass parameter was fixed by the scale  symmetry breaking 
effect, i.e., non-zero value of divergence of dilatation current. In our case,   
dilaton mass is considered as free parameter and that can be justified when the tentative 
dark matter contributions are included, for instance.}.    
This point should be even more  
evident for the radion in the Randall-Sundrum scenario, since the existence of the 
radion $\phi$  is independent of EWSB, and thus it should couple to the $T_{~\mu}^\mu$ 
of the SM fields with the full SM gauge symmetry $G_{\rm SM}$, Eq. (3), and not to 
the form with the unbroken subgroup $H_{\rm SM}$ of the SM.

It is the purpose of this paper to analyze the Higgs-dilaton system using the dilaton 
couplings to the SM fields which respects the full SM gauge interactions, and compare
the results with the most recent LHC data on the Higgs boson.   In Sec.~II, we derive 
the effective Lagrangian for dilaton coupled to the SM fields, and derive the interactions
between them.  Then we perform phenomenological analysis in Sec.~III, comparing 
theoretical predictions based on Eq.~(3) with the LHC data on the Higgs boson, and
derive the constraints on the mass of the 2nd scalar boson and the mixing angle, 
as well as the deviations of quartic and triple couplings of the Higgs bosons.
The results are summarized in Sec.~IV , and the $\beta$ functions for dimensionless 
couplings in the SM are collected in Appendix for convenience.  

\section{Model for the Higgs-dilaton(radion) system}
\subsection{Model Lagrangian}
Let us assume that there is a scale invariant system where scale symmetry is 
spontaneously broken at some high energy scale $f_\phi$, with the resulting 
Nambu-Goldstone boson which is called dilaton $\phi$. 
In terms of $\chi (x) \equiv {\rm e}^{\phi (x) / f_\phi}$, the Lagrangian for the SM 
plus a dilaton  would be written as 
\begin{widetext}
\begin{eqnarray}
{\cal L} 
= &&{\cal L}_{\rm SM} ( \mu^2_H = 0)  
+  \frac{1}{2} f_\phi^2 \partial_\mu \chi \partial^\mu \chi 
- \mu^2_H \chi^2 H^\dagger H  
- \frac{f_\phi^2 m_\phi^2}{4} \chi^4 \left\{ \log \chi - \frac{1}{4}  \right\},
\nonumber \\ 
 &-& \log \left( \frac{\chi}{S(x)}  \right) 
     \left\{ 
          \frac{\beta_{g_1} (g_1)}{2 g_1} B_{\mu\nu} B^{ \mu\nu} 
        + \frac{\beta_{g_2} (g_2)}{2 g_2} W_{\mu\nu}^i W^{i \mu\nu} 
        + \frac{\beta_{g_3} (g_3)}{2 g_3} G_{\mu\nu}^a G^{a \mu\nu} 
     \right\} \nonumber \\
 &+& \log \left( \frac{\chi}{S(x)}  \right) 
     \left\{ 
          \beta_u \left(\bf{Y_u}\right) \bar{Q}_L\tilde{H}u_R 
        + \beta_d \left(\bf{Y_u}\right) \bar{Q}_L{H}d_R 
        + \beta_l \left(\bf{Y_u}\right) \bar{l}_L{H}e_R  + H.c.
     \right\} \nonumber \\
 &+& \log \left( \frac{\chi}{S(x)}  \right) \frac{\beta_\lambda (\lambda) }{4}
     \left(H H^\dagger \right)^2  
\end{eqnarray}
\end{widetext} 
where $S(x)$ is the conformal compensator, which is put to 1 at the end of 
calculation.  Keeping the linear term in $\phi$, we recover the Eq.~(1) with 
$T^\mu_{~\mu}$ being given by Eq.~(3).
Note that the dilaton coupling to the SM fields in this work is  
different from other works in the literature.  In most works, the dilaton is assumed to 
couple to the SM fields in the broken phase with unbroken local    
$SU(3)_c \times U(1)_{\rm em}$ symmetry. 
However if scale symmetry breaking occurs at high energy scale, it would be more 
reasonable to assume that the dilaton couple to the SM Lagrangian as given in the above 
form with the full SM gauge symmetry $SU(3)_c \times SU(2)_L \times U(1)_Y$ imposed. 

The ground state of the potential for the classical Lagrangian is given by either 
$\langle H \rangle = 0 , \langle \chi \rangle = 1$
for the unbroken EW phase, and 
$\langle H \rangle = (0, v/\sqrt{2})^T , \langle \phi \rangle = \bar{\phi}$ for EWSB 
into $U(1)_{\rm em}$,  ignoring the contributions from the vacuum expectation values 
of the scale anomaly, such as $\langle G_{\mu\nu}^a G^{a\mu\nu}\rangle$ etc.

The vanishing tadpole conditions for the correct vacua are given by 
\begin{eqnarray}
\lambda v^2 & = & \mu^2 {\rm e}^{2 \frac{\bar{\phi}}{f_\phi}}, 
\\
\mu^2 v^2 & = & f_\phi m_\phi^2 \bar{\phi}~{\rm e}^{2 \frac{\bar{\phi}}{f_\phi}}. 
\end{eqnarray}
We have used the $\mu^2=-\mu_H^2 > 0$, for convenience.
From these two conditions, one can derive 
\begin{equation}
v^2 = \frac{\mu^2}{\lambda}  {\rm e}^{2 \bar{\phi}/f_\phi} ~~~
{\rm or}~~~\mu^4 = \lambda \bar{\phi} f_\phi m_\phi^2 \ ,
\end{equation}
which solves for $\bar{\phi}$ for given $\mu^2 , \lambda , f_\phi$ and 
$m_\phi^2$.  Note that the Higgs VEV $v$ is fixed by the weak gauge boson
masses $m_W$ and $m_Z$ to be $246$ GeV. 

We will consider the EWSB vacuum, and calculate the (mass)$^2$
matrix for the field fluctuation around the VEV: $ H = (0, (v+h(x))/\sqrt{2})^T$ 
and $\bar{\phi} + \phi$. Note that rescaling of the quantum fluctuation $\phi$ around $\bar{\phi}$
 is necessary, i.e. $\phi~{\rm e}^{\frac{\bar{\phi}}{f_\phi}} \rightarrow  \phi$. After rescaling 
the mass matrix should be
\begin{eqnarray}
{\cal M}^2 (h,\phi) & = &  
\left( \begin{array}{cc}
       m^2_{hh} & m^2_{h\phi} \\
       m^2_{\phi h} & m^2_{\phi\phi} 
       \end{array} \right) 
= \left(  \begin{array}{cc} 
          2 \lambda v^2 & - 2 \frac{\lambda v^3}{f_\phi} {\rm e}^{-2\frac{\bar{\phi}}{f_\phi}}  \\
        - 2 \frac{\lambda v^3}{f_\phi} {\rm e}^{-2\frac{\bar{\phi}}{f_\phi}} & 
          m_\phi^2 {\rm e}^{2 \frac{\bar{\phi}}{f_\phi}} \left( 1 + 2\frac{\bar{\phi}}{f_\phi} \right) 
         \end{array}   \right)   \nonumber \\
& \equiv & \left(  \begin{array}{cc} 
          m_h^2 & - m_h^2 \frac{v}{f_\phi} {\rm e}^{-2\frac{\bar{\phi}}{f_\phi}}  \\
        - m_h^2 \frac{v}{f_\phi} {\rm e}^{-2\frac{\bar{\phi}}{f_\phi}} & 
          \tilde{m}_\phi^2 {\rm e}^{2 \frac{\bar{\phi}}{f_\phi}} 
         \end{array}   \right) ,
\end{eqnarray}
where we define 
\begin{eqnarray}
\tilde{m}_\phi^2 = m_\phi^2 \left( 1 + 2\frac{\bar{\phi}}{f_\phi} \right).
\end{eqnarray}

One can diagonalize this matrix by introducing two mass eigenstates $H_1$ and $H_2$ 
and the mixing angle $\alpha$ between the two states, 
with the following transformation: 
\begin{widetext}
\begin{eqnarray}
m_{H_{1,2}}^2 & = &
\frac{m_h^2+\tilde{m}_\phi^2 {\rm e}^{2\frac{\bar{\phi}}{f_\phi}} 
   \mp 
\sqrt{\left( m_h^2-\tilde{m}_\phi^2{\rm e}^{2\frac{\bar{\phi}}{f_\phi}}\right)^2
             +4{\rm e}^{-4\frac{\bar{\phi}}{f_\phi}} 
                                         \frac{v^2}{f_\phi^2} m_h^4}}{2} \ ,
                                         \\
\tan  \alpha &= & 
\frac{- m_h^2 \frac{v}{f_\phi} {\rm e}^{-2\frac{\bar{\phi}}{f_\phi}}}
     {\tilde{m}_\phi^2 {\rm e}^{\frac{2\bar{\phi}}{f_\phi}} -m_{H_1}^2} \ .
\end{eqnarray}
\end{widetext}
Here we use the basis  
\begin{equation}
\left( \begin{array}{c}
         H_1 \\ H_2 
         \end{array}  \right) 
= \left( \begin{array}{cc}
            \cos \alpha & - \sin \alpha  \\
            \sin \alpha & \cos \alpha 
            \end{array} \right) 
\left( \begin{array}{c}
         h \\ \phi 
         \end{array}  \right).
\end{equation}
Now the interaction Lagrangian between dilaton and the SM fields can be derived
in terms of  $H_1$ and $H_2$.

\subsection{Interaction Lagrangian for dilaton(radion) and the SM Fields}

In this subsection, we derive the interaction Lagrangian between the dilaton(radion) 
and the SM fields both  in the interaction and in the mass eigenstate basis. 

Let us first discuss the interactions of the dilaton(radion) with the SM fermions 
and the SM Higgs boson with the full $G_{\rm SM}$: 
\begin{equation}
{\cal L} (f,\bar{f}, H_{i=1,2} )   =  - \frac{m_f}{v} \overline{f} f h 
= - \frac{m_f}{v} \overline{f} f ( H_1 c_\alpha + H_2 s_\alpha ) ,
\end{equation}
with $s_\alpha \equiv \sin \alpha$ and $c_\alpha = \cos \alpha$. 
The first equality is in the interaction basis, whereas the second one is in
the mass basis.  Note that there is no direct coupling of the dilaton(radion) ($\phi$) 
to the SM chiral fermion at the classical level, namely when we ignore the quantum 
scale anomaly of Yukawa interactions. This is because we have imposed the full SM 
gauge symmetry, Eq. (3).   On the other hand, earlier literature uses the following  
dilaton couplings to the SM fermions assuming the unbroken subgroup $H_{\rm SM} = 
SU(3)_C \times U(1)_Y$: 
\begin{equation}
{\cal L} (f,\bar{f}, \phi )  = - \frac{m_f}{f_\phi} \bar{f}f 
          \phi~{\rm e}^{-\bar{\phi}/f_\phi} .
\end{equation}
Note that there is no proper limit where the earlier result (14) based on 
$T_{~\mu}^\mu$ with unbroken subgroup of the SM gauge symmetry 
$H_{\rm SM} = SU(3)_C \times U(1)_{\rm em}$ approaches our result 
(13) based on $T_{~\mu}^\mu$ with the full SM gauge symmetry 
$G_{\rm SM} = SU(3)_C \times SU(2)_L \times U(1)_Y$.  
This shows that it is very important to impose which gauge symmetry on the 
fundamental Lagrangian. It should be the full SM gauge symmetry 
$G_{\rm SM} = SU(3)_C \times SU(2)_L \times U(1)_Y$ rather than its unbroken subgroup
$H_{\rm SM} = SU (3)_C \times U(1)_{\rm em}$ that has been widely used in earlier 
literature, when we consider new physics at EW scale and the new physics scale is not 
known  \footnote{
Similar observation was made in Ref.~\cite{Baek:2011aa} in the context of singlet fermion 
dark matter model with Higgs portal. 
There it was shown that the effective Lagrangian approach gives us completely wrong 
answer in direct detection cross section for dark matter and nucleon and Higgs boson 
phenomenology.  The same conclusion would apply even if one assumes that the interaction
between the SM Higgs boson $h$ and the fermion DM $\psi$ is given by dim-4 operator 
$h \bar{\psi} \psi $, which in fact respects only the unbroken SM gauge group $H_{\rm SM}$ 
and not the full gauge symmetry $G_{\rm SM}$. When we construct the renormalizable 
model with the full SM gauge symmetry, we would recover the model presented in 
Ref.~\cite{Baek:2011aa}, and the model based on $H_{\rm SM}$ would give wrong physics. 
More detail will be described elsewhere.}.

The same argument applies to other interactions of the dilaton(radion) with 
the SM gauge bosons or the SM Higgs boson.  We list them below for completeness: 
\begin{eqnarray}
{\cal L} (g,g,H_{i=1,2}) & = & 
- \frac{ {\rm e}^{-\bar{\phi} /f_\phi}}{f_\phi}~\frac{\beta_3 (g_3)}{2 g_3} 
 G_{\mu\nu} G^{\mu\nu} \phi
\nonumber \\
& = & 
 -\frac{ {\rm e}^{-\bar{\phi} /f_\phi}}{f_\phi}~\frac{\beta_3 (g_3)}{2 g_3} 
 G_{\mu\nu} G^{\mu\nu}  ( -H_1 s_\alpha + H_2 c_\alpha ).
\\
{\cal L} (W,W,H_{i=1,2}) & = & \frac{2 m_W^2 }{v} 
W_\mu^+ W^{- \mu}  h 
- \frac{ {\rm e}^{-\bar{\phi} /f_\phi}}{f_\phi}~\frac{\beta_2 (g_2)}{2 g_2} 
 W_{\mu\nu} W^{\mu\nu} \phi
\nonumber \\
& = & \frac{2 m_W^2 }{v} 
W_\mu^+ W^{- \mu}  \left( H_1  c_\alpha + H_2  s_\alpha \right)  \nonumber  \\
&-&  \frac{ {\rm e}^{-\bar{\phi} /f_\phi}}{f_\phi}~\frac{\beta_2 (g_2)}{2 g_2} 
 W_{\mu\nu} W^{\mu\nu}  ( -H_1 s_\alpha + H_2 c_\alpha ) .
\\
{\cal L} (Z,Z,H_{i=1,2}) & = & \frac{m_Z^2 }{v} 
Z_\mu Z^{ \mu}  h 
- \frac{ {\rm e}^{-\bar{\phi} /f_\phi}}{f_\phi}~\left\{c_W^2\frac{\beta_2 (g_2)}{2 g_2} 
+s_W^2 \frac{\beta_1 (g_1)}{2 g_1}\right\}Z_{\mu\nu} Z^{\mu\nu} \phi
\nonumber \\
& = & \frac{m_Z^2 }{v} 
Z_\mu Z^{ \mu}  \left( H_1  c_\alpha + H_2 s_\alpha \right)   \nonumber  \\
&-& \frac{ {\rm e}^{-\bar{\phi} /f_\phi}}{f_\phi}~\left\{c_W^2 \frac{\beta_2 (g_2)}{2 g_2} 
+s_W^2 \frac{\beta_1 (g_1)}{2 g_1}\right\}Z_{\mu\nu} Z^{\mu\nu}  ( -H_1 s_\alpha + H_2 c_\alpha ).
\\ 
{\cal L} (\gamma,\gamma,H_{i=1,2}) & = & 
-\frac{ {\rm e}^{-\bar{\phi} /f_\phi}}{f_\phi}~\left\{s_W^2\frac{\beta_2 (g_2)}{2 g_2} 
+c_W^2\frac{\beta_1 (g_1)}{2 g_1} \right\}
 F_{\mu\nu} F^{\mu\nu}\phi \nonumber \\
& = &-\frac{ {\rm e}^{-\bar{\phi} /f_\phi}}{f_\phi}~\left\{s_W^2\frac{\beta_2 (g_2)}{2 g_2} 
+c_W^2\frac{\beta_1 (g_1)}{2 g_1} \right\} 
 F_{\mu\nu} F^{\mu\nu}  ( -H_1 s_\alpha + H_2 c_\alpha ).
\\
{\cal L} (\gamma,Z,H_{i=1,2}) & = & 
-\frac{ {\rm e}^{-\bar{\phi} /f_\phi}}{f_\phi}~2s_W c_W\left\{\frac{\beta_2 (g_2)}{2 g_2} 
-\frac{\beta_1 (g_1)}{2 g_1} \right\}
 Z_{\mu\nu} F^{\mu\nu}\phi \nonumber \\
&=&-\frac{ {\rm e}^{-\bar{\phi} /f_\phi}}{f_\phi}~2s_W c_W\left\{\frac{\beta_2 (g_2)}{2 g_2} 
-\frac{\beta_1 (g_1)}{2 g_1} \right\}
 Z_{\mu\nu} F^{\mu\nu}( -H_1 s_\alpha + H_2 c_\alpha ). 
\end{eqnarray}
The $\beta$ functions for the SM gauge groups are listed in the Appendix A for 
convenience.  The SM Higgs field $h$ will interact with gluons or photons  
just as in the standard model case, and we have to add these to the above interaction
Lagrangian.

The offshoot of our approach is that the dilaton $\phi$ mixes with the SM Higgs boson 
$h$, and couples to the SM fields through quantum scale anomaly in addition to  the 
classical scale symmetry breaking term, i.e. $\mu^2_H H^\dagger H$.
Since the dilaton $\phi$ and the SM Higgs boson $h$ mix with each other to make 
two scalar bosons $H_1$ and $H_2$, their couplings to the SM fermions will be reduced  
by a universal amount due to the mixing effects~\cite{choi_jung_ko},  
while their couplings to the SM gauge bosons, especially to gluons,  could be further 
modified by quantum scale anomalies. 
This observation has a very tantalizing implication for Higgs signals at the LHC, 
which will be elaborated in the following.  Since there are two scalar bosons,   
we take one of them to be 126~GeV resonance that was observed recently at the LHC. 
Since the dilaton(radion) $\phi$  couplings to the trace anomaly of the SM fields (Eq.~(3))
are distinctly different from the interactions between the SM fields and other singlet 
scalar  bosons appearing in various extensions of the SM~\cite{choi_jung_ko},  
phenomenological consequences of the Higgs-dilaton mixing are analyzed separately 
in this paper.

\section{Implications on the LHC Higgs data}
\subsection{Analysis Strategy}
Compared with the SM Higgs boson, the Higgs-dilaton system considered in this paper 
has only two more parameters ($m_{\phi}$ and $f_\phi$), which makes phenomenological 
analysis feasible.   Two scalars $\phi$ and $h$ mix with each other after EWSB,  
leading to two mass eigenstates $H_1$ and $H_2$.   Fixing one Higgs boson mass to be 
$126$~GeV, all other parameters in the Lagrangian such as the other Higgs boson mass, 
the mixing angle $\alpha$,  triple and quartic couplings of $H_1$ and $H_2$, are all 
expressed as functions of $m_\phi$ and $f_\phi$.  Likewise, their decay widths and 
branching ratios are completely fixed as functions of $m_\phi$ and $f_\phi$. 
In the numerical analysis, we assume $f_\phi \geq v$, following our spirit that the
spontaneous scale symmetry breaking scale occurs before electroweak symmetry breaking. 

As mentioned in the previous section, the interactions of the Higgs boson to the SM 
particles are modified in two different ways compared with the SM, via mixing with 
dilaton and the quantum scale anomalies.  Note that the modification due to quantum 
scale anomalies are very small that their effects are negligible in most cases, 
except for the gluon-gluon and $\gamma\gamma$ couplings to the Higgs 
boson through scale anomaly associated with $SU(3)_C \times U(1)_{\rm em}$ 
gauge interaction.  Therefore the branching ratios of 
physical Higgs bosons decaying into the SM fermions are suppressed relative to those 
of the SM Higgs boson by mixing angle, whereas those into the SM gauge bosons could 
be  modified through quantum scale anomaly.

For a given $( m_\phi , f_\phi )$, we  calculate the signal strength of each scalar boson 
into a specific final state:
\begin{equation}
\mu_i (f.s.) = \frac{\sigma_H (production) \times B (H_i \rightarrow f.s.) }{
\sigma_H (production)_{SM} \times B (H_i \rightarrow f.s.)_{SM} },
\end{equation}
where `$f.s.$' means a specific `final states', $WW^*, ZZ^*, \gamma \gamma, f\bar{f}$, etc.
The subscript $i =1,2$ represents two scalar bosons in the mass eigenstates, and 
the `production' denotes the production mechanisms such as gluon-gluon fusion (ggF), 
vector boson fusion (VBF),   Higgs production associated with vector boson (VH), 
and top quark pair production associated with Higgs boson (ttH). 


\begin{table}[t]
\begin{center}
\begin{tabular}{|c|c|c|}
\hline
Decay & Production & $\mu_i$ \\
\hline
\hline
       & $Combined$ & \parbox{5cm}{ ~ \\ ATLAS\cite{ATLAS13-012} : $1.65^{-0.3}_{+0.35}$ \\ 
                                 CMS\cite{CMS13-001}  : $0.78^{-0.26}_{+0.28}$ \\    } \\ 
\cline{2-3}
$\gamma\gamma$ & $ggF$ & \parbox{5cm}{ ~\\ ATLAS\cite{ATLAS13-012} : $1.6^{-0.36}_{+0.42}$ \\ } \\ 
\cline{2-3}
       & $VBF$ & \parbox{5cm}{ ~ \\ ATLAS\cite{ATLAS13-012} : $1.7^{-0.89}_{+0.94}$ \\ } \\ 
\hline
\hline
\hline
       & $Combined$ & \parbox{5cm}{ ~ \\ ATLAS\cite{ATLAS13-013} : $1.7^{-0.4}_{+0.5}$ \\ 
                                 CMS\cite{Chatrchyan:2013mxa}  : $0.93^{-0.25}_{+0.29}$ \\    } \\ 
\cline{2-3}
$ZZ^*$ & $ggF$ & \parbox{5cm}{ ~\\ ATLAS\cite{ATLAS13-013} : $1.8^{-0.5}_{+0.8}$ \\
                                     CMS\cite{Chatrchyan:2013mxa} : $0.8^{-0.36}_{+0.46}$ \\    } \\ 
\cline{2-3}
       & $VBF(VH)$ & \parbox{5cm}{ ~ \\ ATLAS\cite{ATLAS13-013} : $1.2^{-1.4}_{+3.8}$ \\
                               CMS\cite{Chatrchyan:2013mxa} : $1.7^{-2.1}_{+2.2}$ \\   } \\ 
\hline
\hline
\hline

$WW^*$ & $Combined$ & \parbox{5cm}{ ~\\ ATLAS\cite{ATLAS13-030} : $1.01^{-0.31}_{+0.31}$ \\
                                 CMS\cite{Chatrchyan:2013iaa}  : $0.72^{-0.18}_{+0.2}$ \\    } \\ 
\hline
\hline
\hline

$bb$ & $VH$ & \parbox{5cm}{ ~\\ ATLAS\cite{ATLAS13-079} : $0.2^{-0.7}_{+0.7}$ \\
                                     CMS\cite{Chatrchyan:2013zna} : $1.0^{-0.5}_{+0.5}$  \\    } \\ 
\hline
\hline
\hline
$\tau\tau$ & $Combined$ & \parbox{5cm}{ ~\\ ATLAS\cite{ATLAS13-108} : $1.4^{-0.4}_{+0.5}$ \\
                                     CMS\cite{CMS13-004} : $1.1^{-0.4}_{+0.4}$ \\    } \\ 
\hline

\end{tabular}
\end{center}
\caption{Signal strengths reported by ATLAS and CMS.}
\end{table}

In case of decays of two physical scalar bosons, the dominant effect of dilaton results 
in the coupling suppression via the mixing between $h$ and $\phi$. 
On the other hand, in their  production parts, there are further modifications on the ggF 
generated by quantum scale anomaly associated with color $SU(3)_C$ gauge fields. 
This kind of modification by scale anomaly is small in other production channels, i.e.,  
VBF, VH and ttH.  Consequently, we can expect that the ggF initiated processes can 
be significantly modified  by quantum scale anomaly but other channels are suppressed 
just by the mixing angle.

\subsection{Confronting the LHC Higgs data and predictions for the Higgs self-couplings 
and the mass of the extra boson}
We perform the analyses for two distinct cases. The first case is that the heavy mode 
$H_2$ is identified as observed 126 GeV boson, with extra light mode. The other case 
is that the light mode $H_1$ is identified as observed 126 GeV boson, with extra heavy 
mode.
The CMS and ATLAS Collaborations reported the results based on five- and seven-different channels, 
respectively~\cite{ATLAS13-012,ATLAS13-013,ATLAS13-030,ATLAS13-079,ATLAS13-108,CMS13-001,Chatrchyan:2013mxa,Chatrchyan:2013iaa,Chatrchyan:2013zna,CMS13-004}.(See Table 1.) 
The most recent CMS results are consistent with the SM even including the diphoton 
decay channel~\cite{CMS13-001},  which was larger than the SM value in the previous 
analysis~\cite{Chatrchyan:2012ufa}. 
The enhancement in diphoton mode is still there in the ATLAS report,  and 
also in the $ZZ^*$ mode with less significance. 

Considering the current situation of conflicting data on $H\rightarrow \gamma\gamma$,
we consider two separate cases reported by CMS and ATLAS Collaborations. 
For each case, we perform the $\chi^2$ analysis and select the parameter sets 
within the $3\sigma$ range around the each $\chi^2$ minimum \cite{avni:1976}. 

\subsubsection{Case I : $126$~{\rm GeV} $H_2$ and extra light $H_1$ : 
$m_{H_1} < m_{H_2}=126~{\rm GeV}$}

\begin{figure*}[t]
\includegraphics[height=7cm,width=8cm]{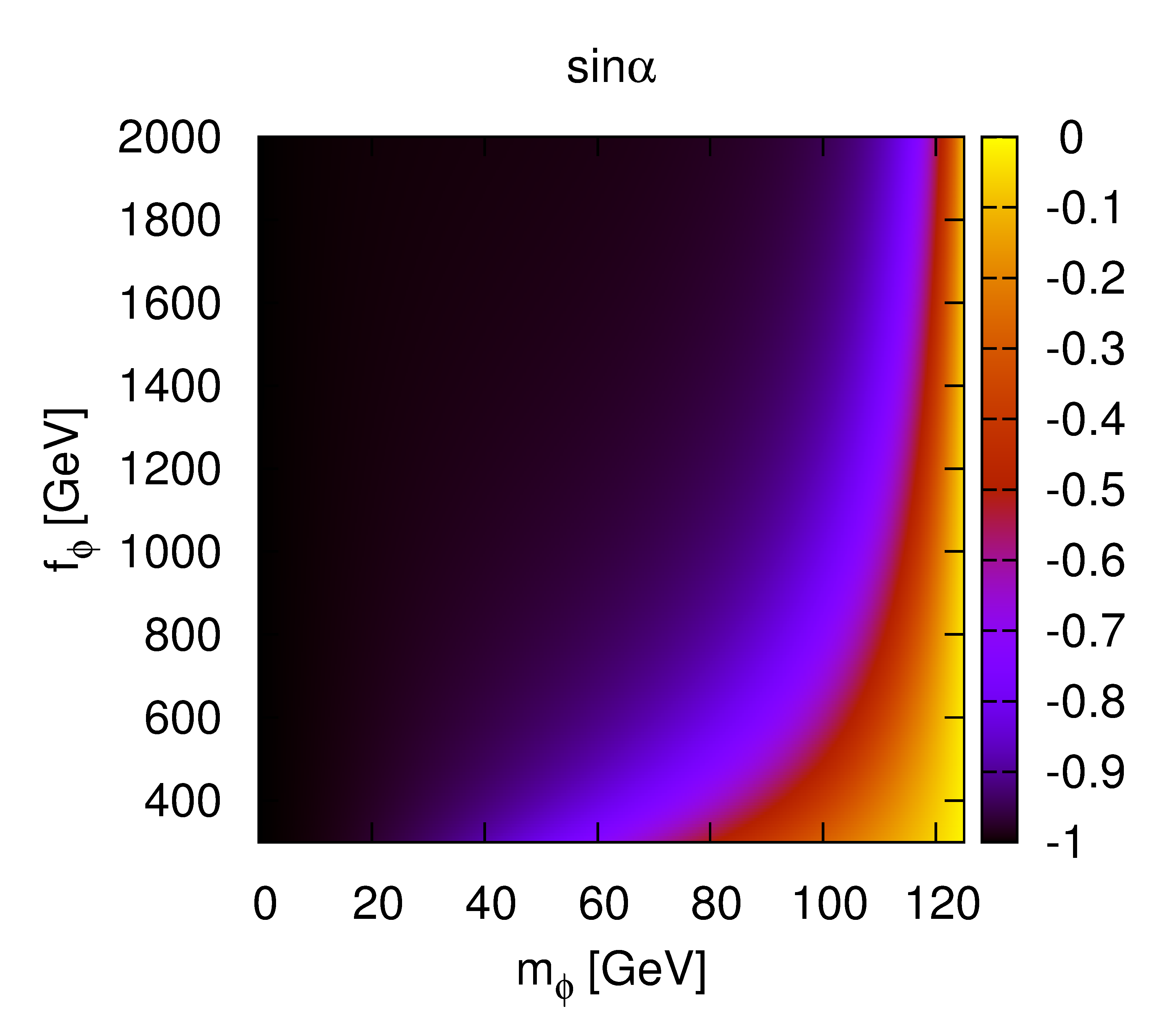}
\includegraphics[height=7cm,width=8cm]{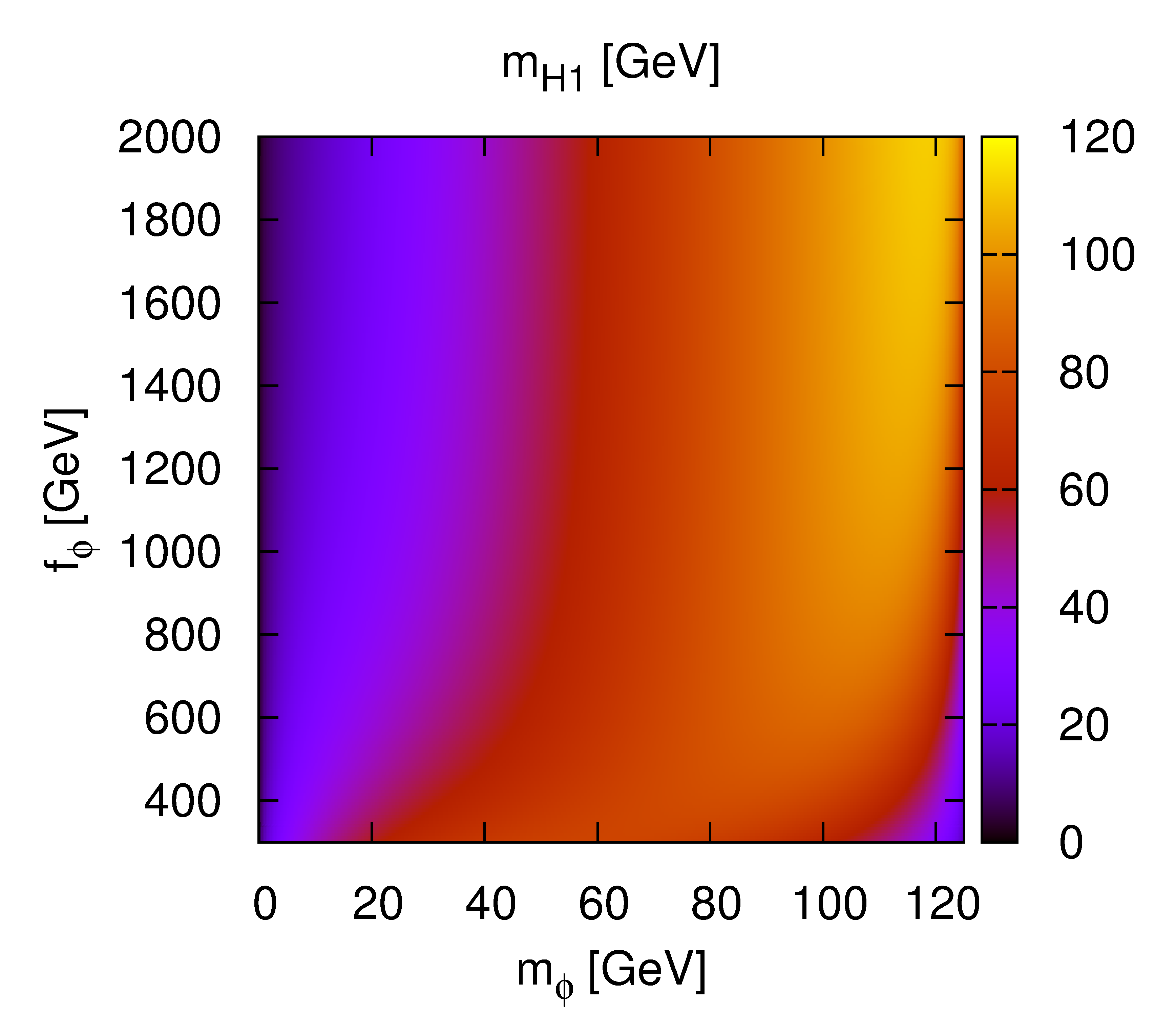}
\caption{Contour plots for the mixing angle and extra light boson mass in the $(m_\phi,f_\phi)$ plane.}
\label{fig:mf-s}
\end{figure*}

Let us start with the case that heavier $H_2$ is the observed $126$~GeV boson. 
All the physical observables are functions of $(m_\phi,f_\phi)$, which can be 
traded with the mixing angle $\alpha$ and the second Higgs mass $m_{H_1}$. 
In Fig.\ref{fig:mf-s}, we show that the mixing angle $\alpha$ and the extra light boson 
mass $m_{H_1}$ are fully determined in the $(m_\phi,f_\phi)$ plane.

With the identification of the observed Higgs boson as $m_{H_2}=126$ GeV, 
we also put the constraints of $3\sigma$ range 
around the minimum $\chi^2$.   In addition to that, we also consider the experimental constraint 
for the light scalar particle 
that is determined by the LEP experiment \cite{Barate:2003sz}.

\begin{figure*}[t]
\includegraphics[height=8cm,width=8cm]{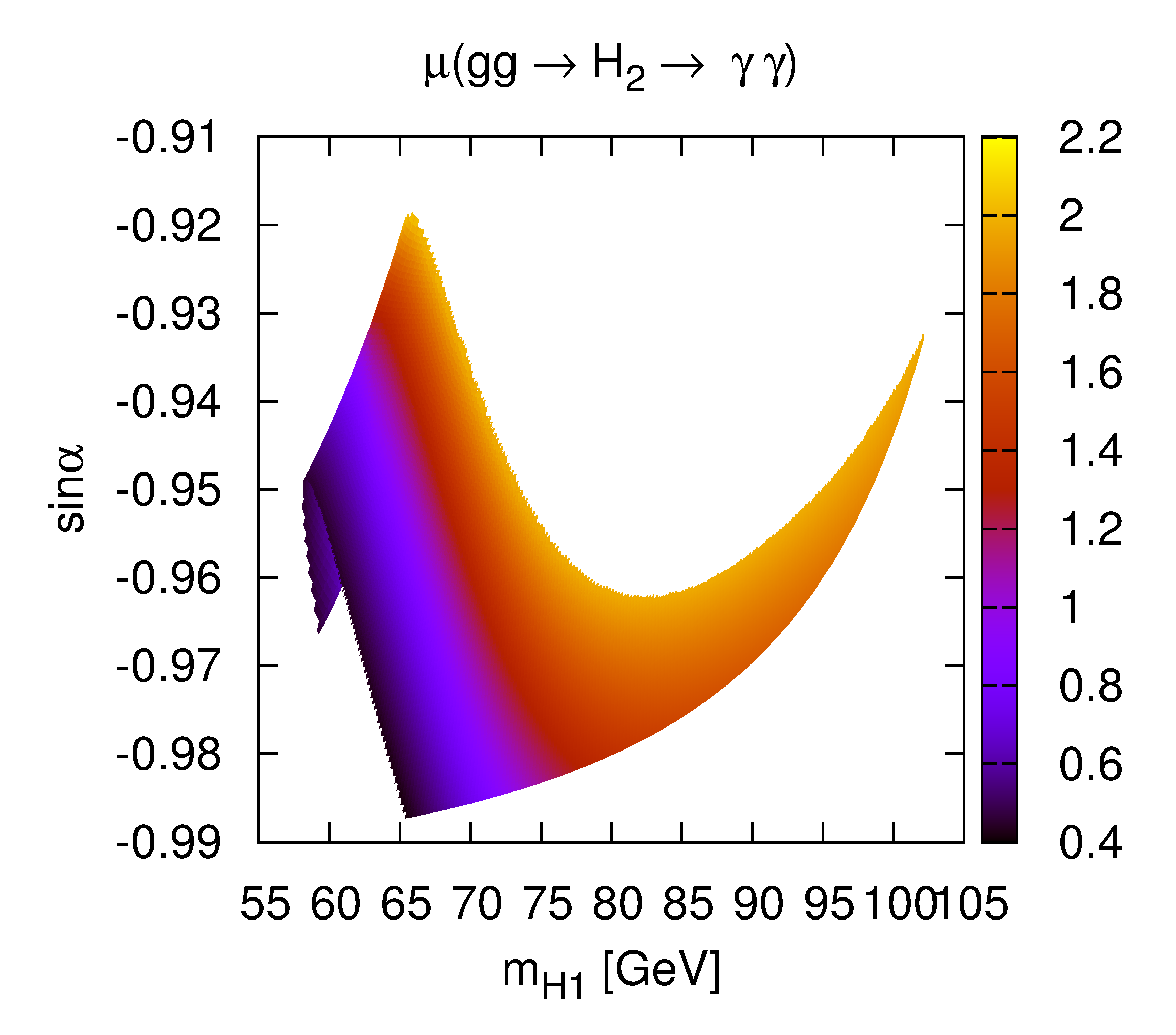}
\includegraphics[height=8cm,width=8cm]{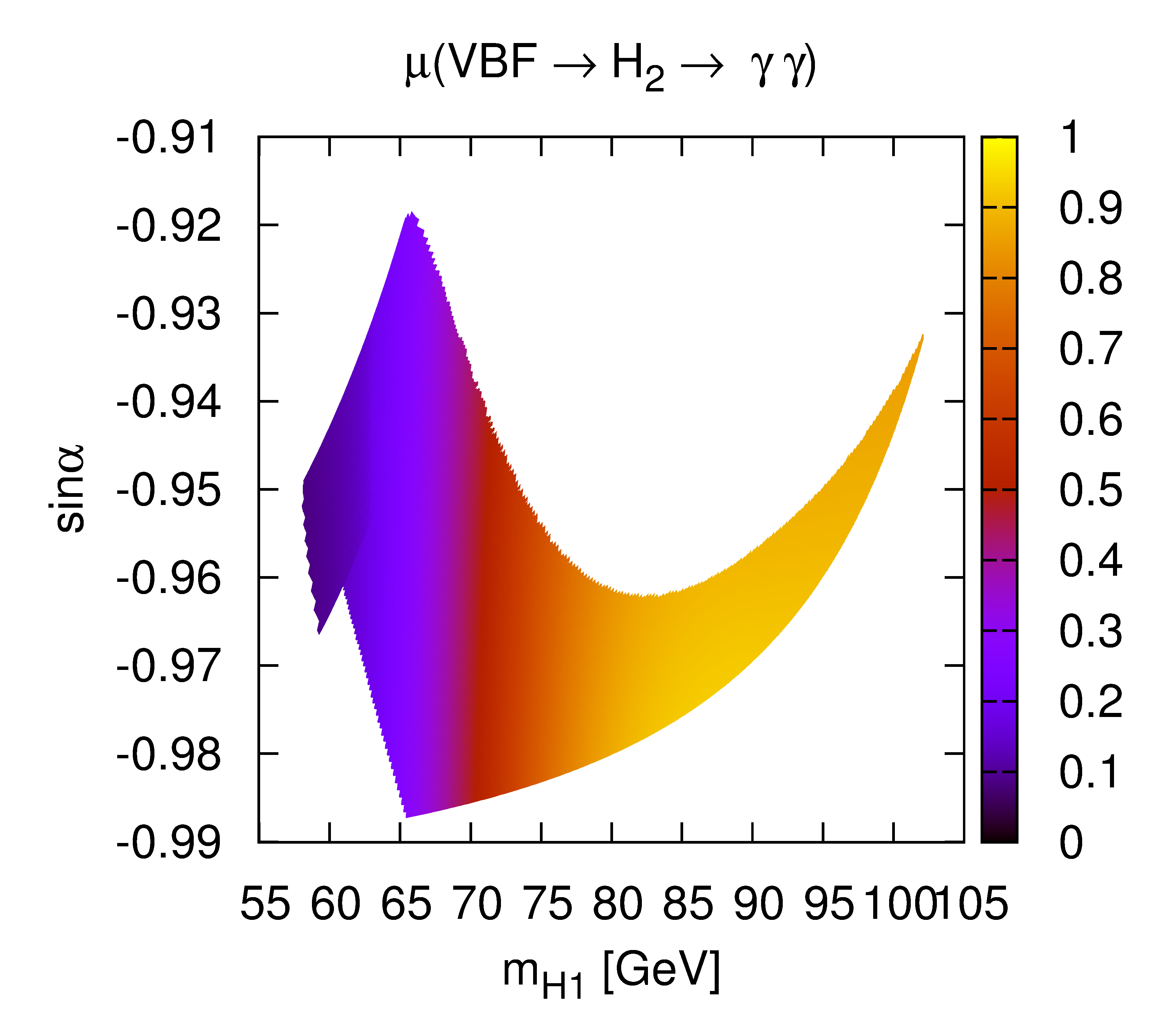}
\caption{Contour plots for the signal strengths for diphoton productions:
 ggF (left) and VBF (right) initiated processes. $H_2$ is fixed as $m_{H_2}=126$ GeV boson.}
\label{fig:s-AA}
\end{figure*}

Considering these three constraints, the allowed parameter region is shown in 
Fig.\ref{fig:s-AA} in the $( m_{H_1} , \sin \alpha )$ plane.  The colored columns denote
the signal strengths.   As noted in Sec.~3.1, the dilaton production from gluon fusion 
($gg\rightarrow \phi$) can be enhanced due to the QCD scale anomaly and  thus 
can compete with the SM Higgs production from gluon fusion ($gg \rightarrow h$).  
Therefore the Higgs signal strengths depend mainly on the production channels 
rather than the decay channels, and we present the $\gamma\gamma$ channel 
only in Fig.~1.  
One can see that the ggF initiated process can be  modified significantly compared to the 
SM value.  On the other hand, the VBF initiated one is suppressed by the mixing 
angle only, so that its signal strength is always smaller than one.  Also note that the 
allowed region for the mixing angle $\alpha$ is highly constrained around 
$\alpha \sim -\frac{\pi}{2}$. This means that the observed $126$~GeV boson is largely 
SM-like and the extra light mode is dilaton-like, namely $H_2 \simeq h$ and 
$H_1 \simeq \phi$.  Even though the mixing angle is close to $-\pi/2$ and $H_2 \simeq h$, 
rather large modification is possible from the mixing with the dilation through the tuning 
of the input parameters $m_\phi$ and $f_\phi$. 
There should be an extra light scalar mode $H_1$ whose mass is constrained to be 
in the range $m_{H_1} \sim [58,104]$~GeV, which is a prediction of our model. 

\begin{figure*}[t]
\includegraphics[height=8cm,width=8cm]{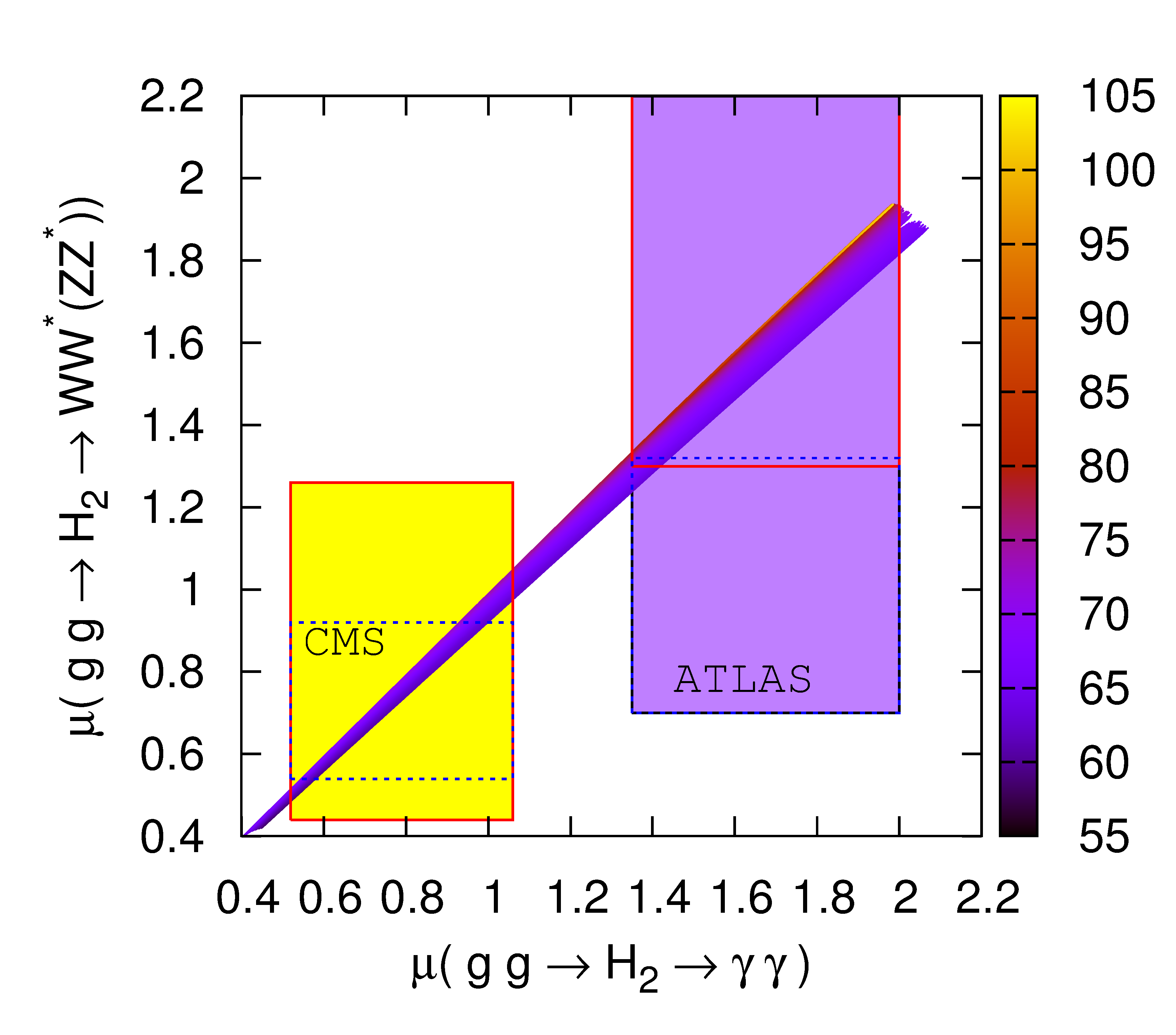}
\includegraphics[height=8cm,width=8cm]{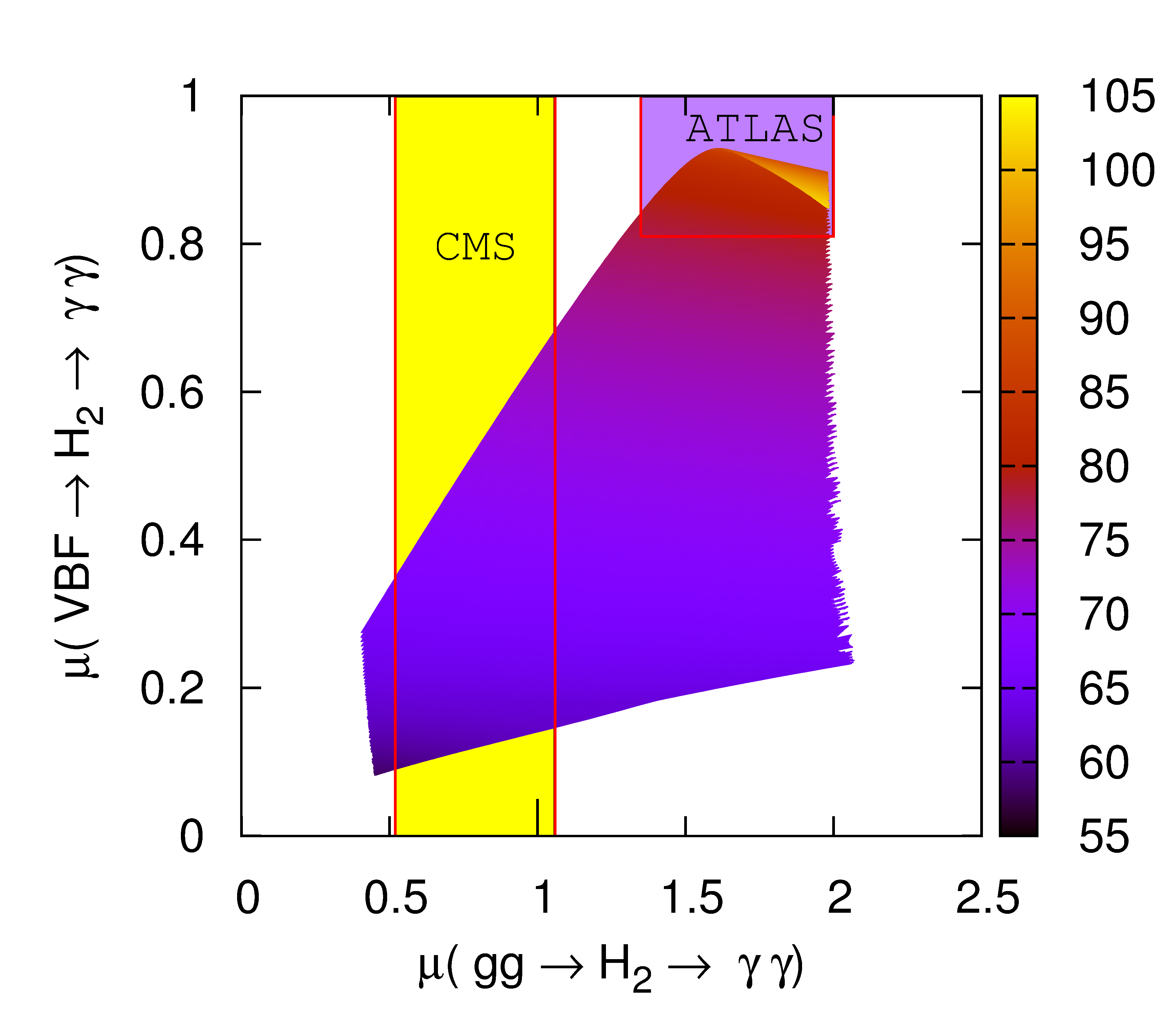}
\caption{Left : Correlation between the signal strengths of ggF initiated $WW^*(ZZ^*)$ and $\gamma\gamma$ production. Blue dotted and 
red solid boxes mean the $1-\sigma$ allowed ranges for $WW^*$ and $ZZ^*$ each.
Right : Correlation between ggF and VBF initiated diphoton production. $H_2$ is fixed as $m_{H_2}=126$ GeV boson.
The colored columns denote the mass of the lighter Higgs $H_1$. }
\label{fig:s-corr}
\end{figure*} 

Since the model has only two more input parameters $(m_\phi, f_\phi)$, 
some observables are highly correlated, which make the generic  
signals of the model.  In Fig.\ref{fig:s-corr}, we show two such correlations with the 
contours of $m_{H_1}$ in different colors. 
The left plot shows a strong correlation between the ggF-initiated signal strengths 
$WW^*(ZZ^*)$ and $\gamma\gamma$ production process.  
The correlation is almost linear, since the scale anomaly contribution to the ggF initiated 
process is dominant.
The slope of the correlation slightly  deviates  from one `1' because 
of the small difference between the $SU(2)_W$ and $U(1)_{em}$  scale anomalies.
The yellow and purple boxes are showing the $1\sigma$ observations by CMS and 
ATLAS. The right plot shows the correlation between the signal strengths of different initial 
states but the same final states, the diphoton channels.
Though the correlation is not that strong as the left plot, the ATLAS data tends to 
prefer the larger value of $m_{H_1}$ 
\footnote{Note that the CMS did not report the results depending on the initial production channels.}.

\begin{figure*}[t]
\includegraphics[height=7cm,width=8cm]{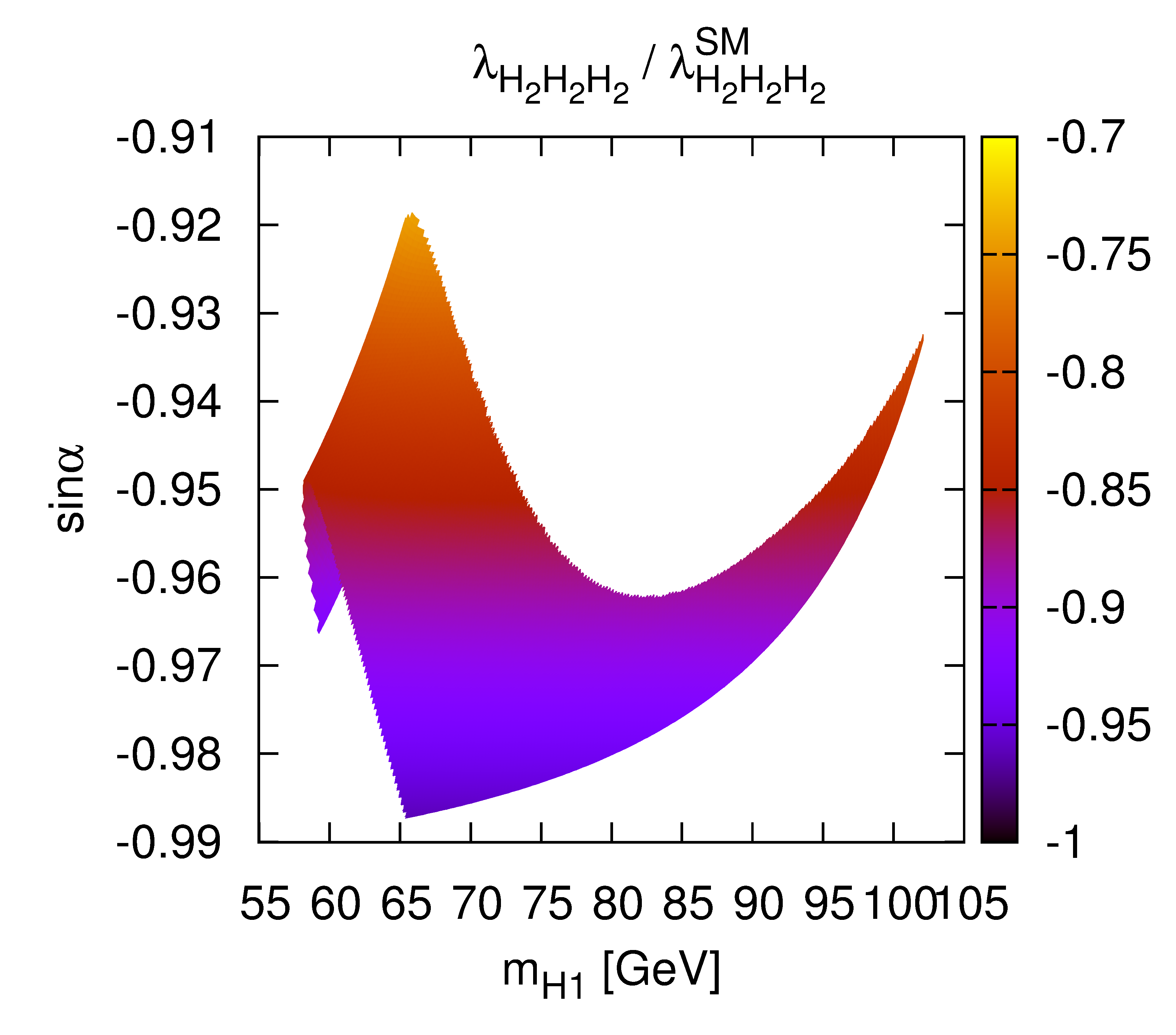}
\includegraphics[height=7cm,width=8cm]{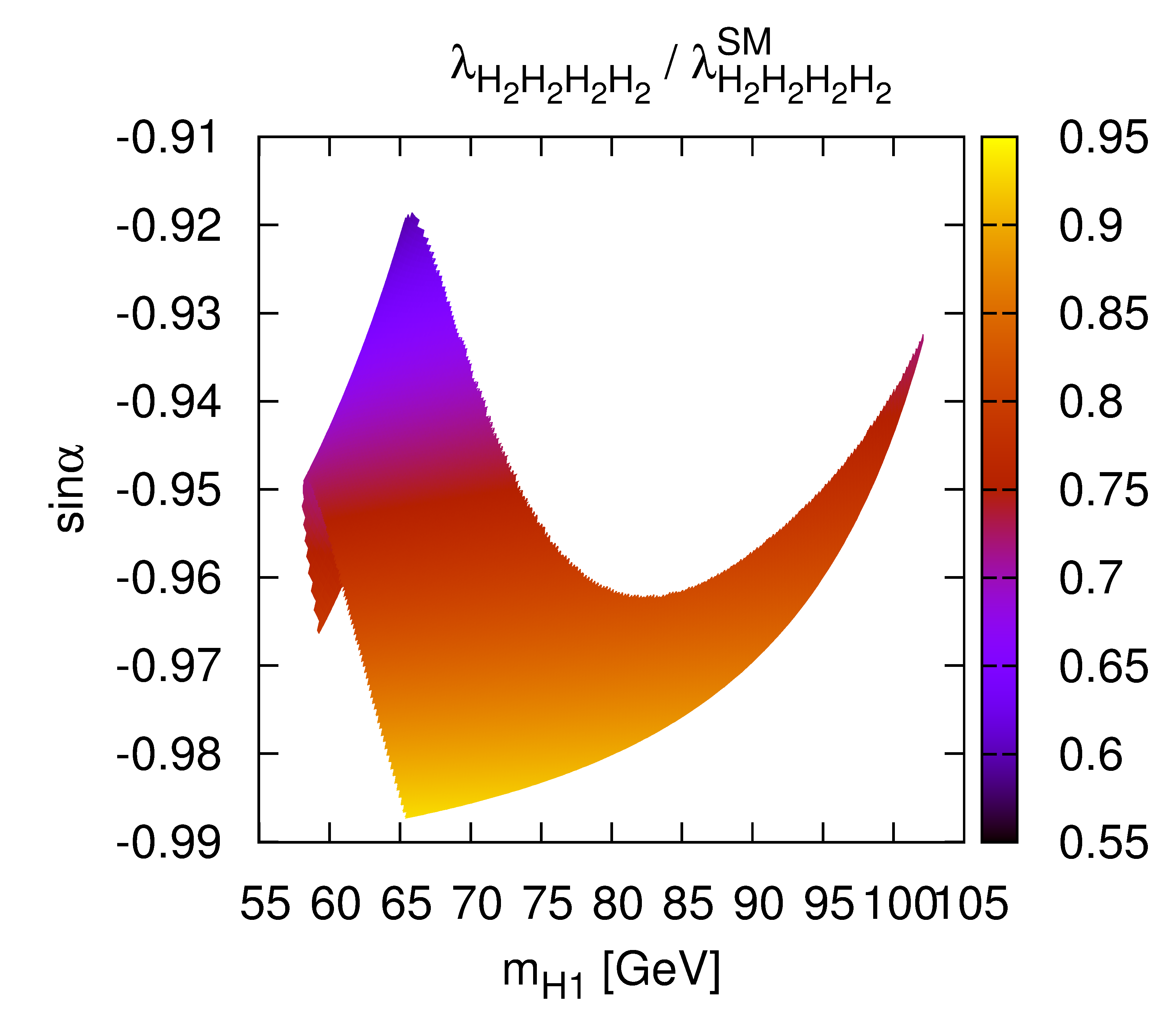}
\caption{Contour plots for triple coupling (left) and quartic coupling (right) for the observed $126$ GeV boson $H_2$,
normalized with the SM values. }
\label{fig:s-34}
\end{figure*} 

\begin{figure*}[t]
\includegraphics[height=7cm,width=8cm]{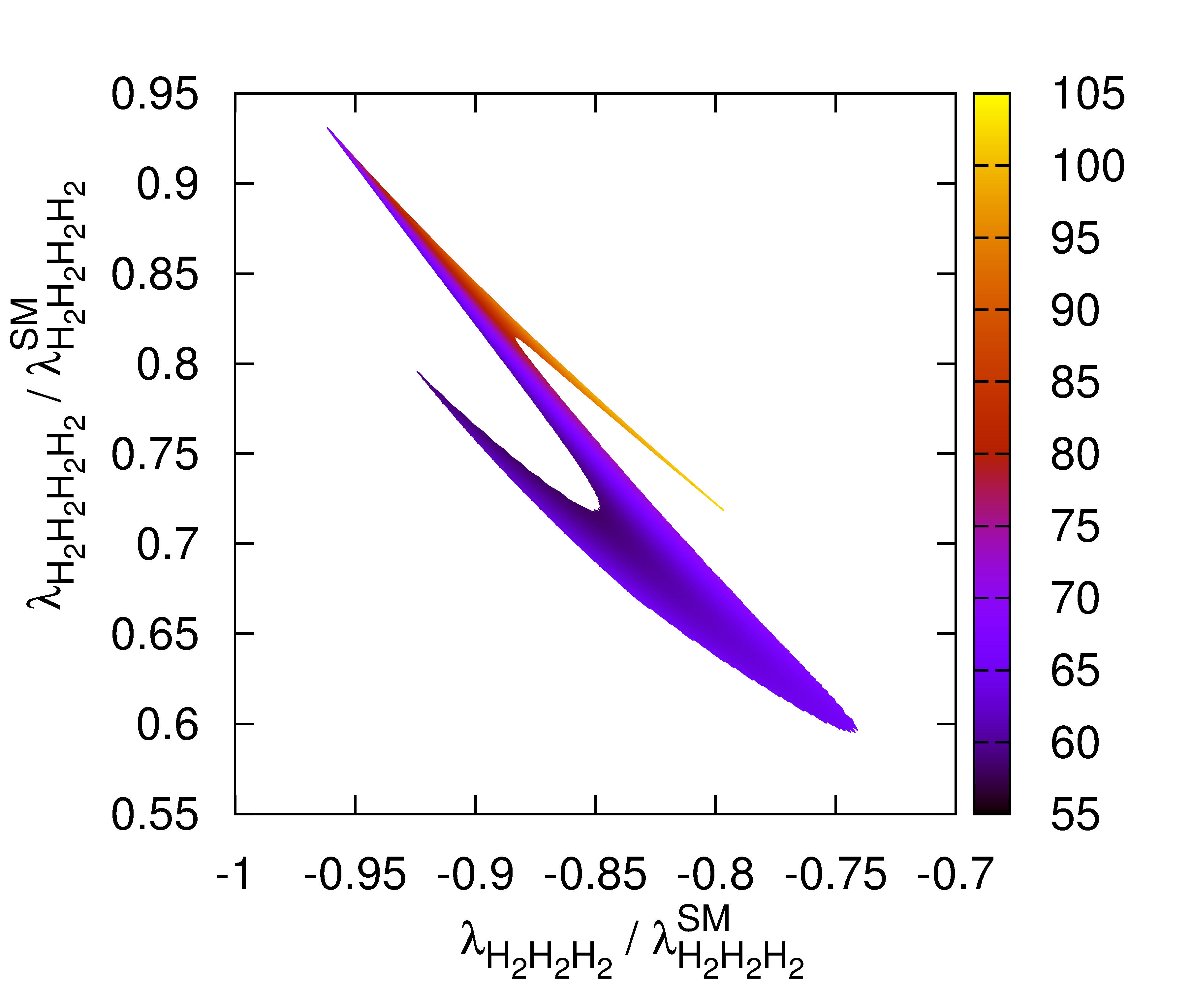}
\caption{Correlation between triple and quartic couplings for the observed $126$ GeV boson $H_2$, 
normalized with the SM values. The colored columns denote the mass of the lighter Higgs $H_1$.}
\label{fig:s-34corr}
\end{figure*} 
Triple and quartic couplings for the $H_2  (m_{H_2}=126~{\rm GeV})$ are completely determined within this model,  making distinct discriminators for this model. 
In the allowed parameter region, the predictions for triple and quartic couplings of the 
SM-like Higgs boson $H_2$ are shown in Fig.\ref{fig:s-34}. One can see that triple and 
quartic couplings are suppressed compared with the SM values, depending on the $H_1$ 
mass. 
Especially for the triple coupling it gives relative minus sign compared 
to the SM value, which would result in the constructive interference between the 
box diagram and the triangle diagram with the $s-$channel $H_2$ propagator, and 
thus increase the $H_2$ pair production in $gg\rightarrow H_2 H_2$ \cite{Plehn:1996wb}.
In addition, we observe a strong correlation between triple and quartic 
couplings, which is presented in Fig.\ref{fig:s-34corr}.   
Along with the $H_1$ mass, the triple and quartic couplings are highly inter-related. 
This will be the strong distinctive signal for testing the model, which could be probed 
at the upcoming LHC run and at the ILC. 

\subsubsection{Case II : $126$~{\rm GeV} $H_1$ and extra heavy $H_2$ : 
$m_{H_1}=126~{\rm GeV} < m_{H_2}$}
\begin{figure*}[t]
\includegraphics[height=7cm,width=8cm]{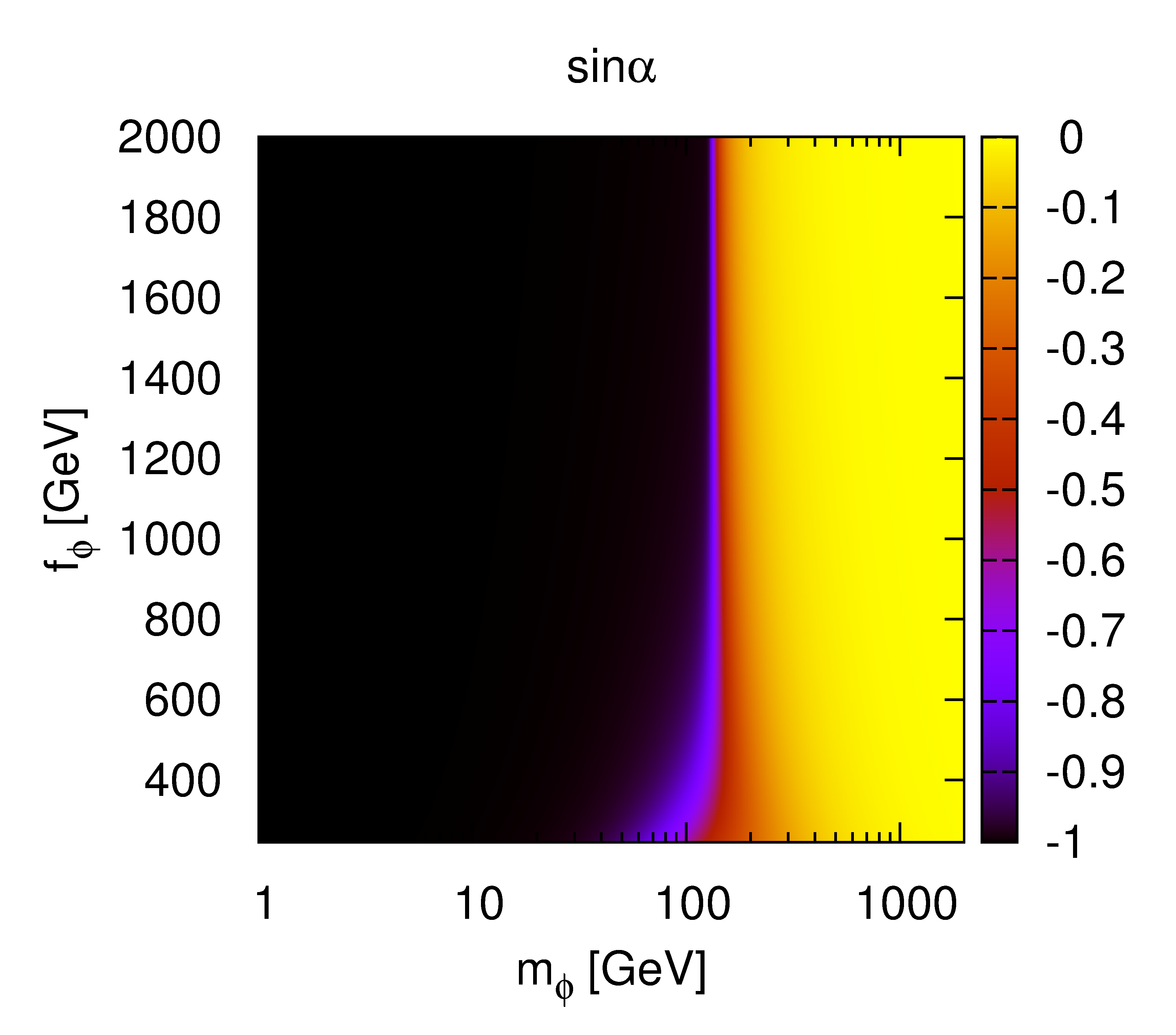}
\includegraphics[height=7cm,width=8cm]{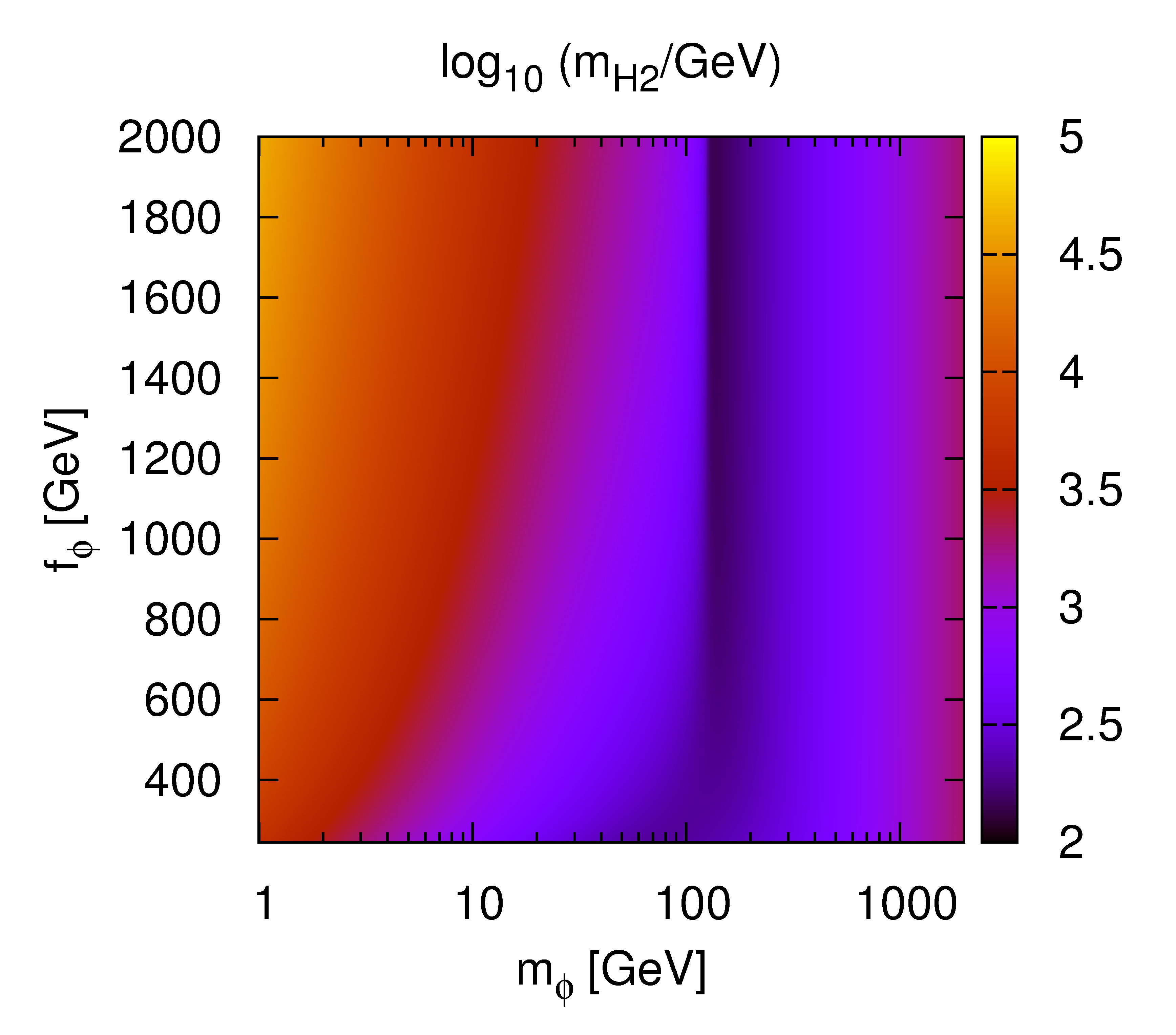}
\caption{Contour plots for the mixing angle and extra heavy boson mass in the $(m_\phi,f_\phi)$ plane. 
Logarithmic scales are used for horizontal axis and extra heavy boson mass for clarity.}
\label{fig:mf-h}
\end{figure*}

Let us move to the other case where the observed $126$~GeV boson is lighter one, 
$H_1$. In this case there is an extra heavy mode named $H_2$. 
As before, we first show  in Fig.~\ref{fig:mf-h} the contour plots 
of the mixing angle $\alpha$ and the extra heavy scalar boson mass $m_{H_2}$ 
in the $(m_\phi,f_\phi)$ plane.

\begin{figure*}[t]
\includegraphics[height=8cm,width=8cm]{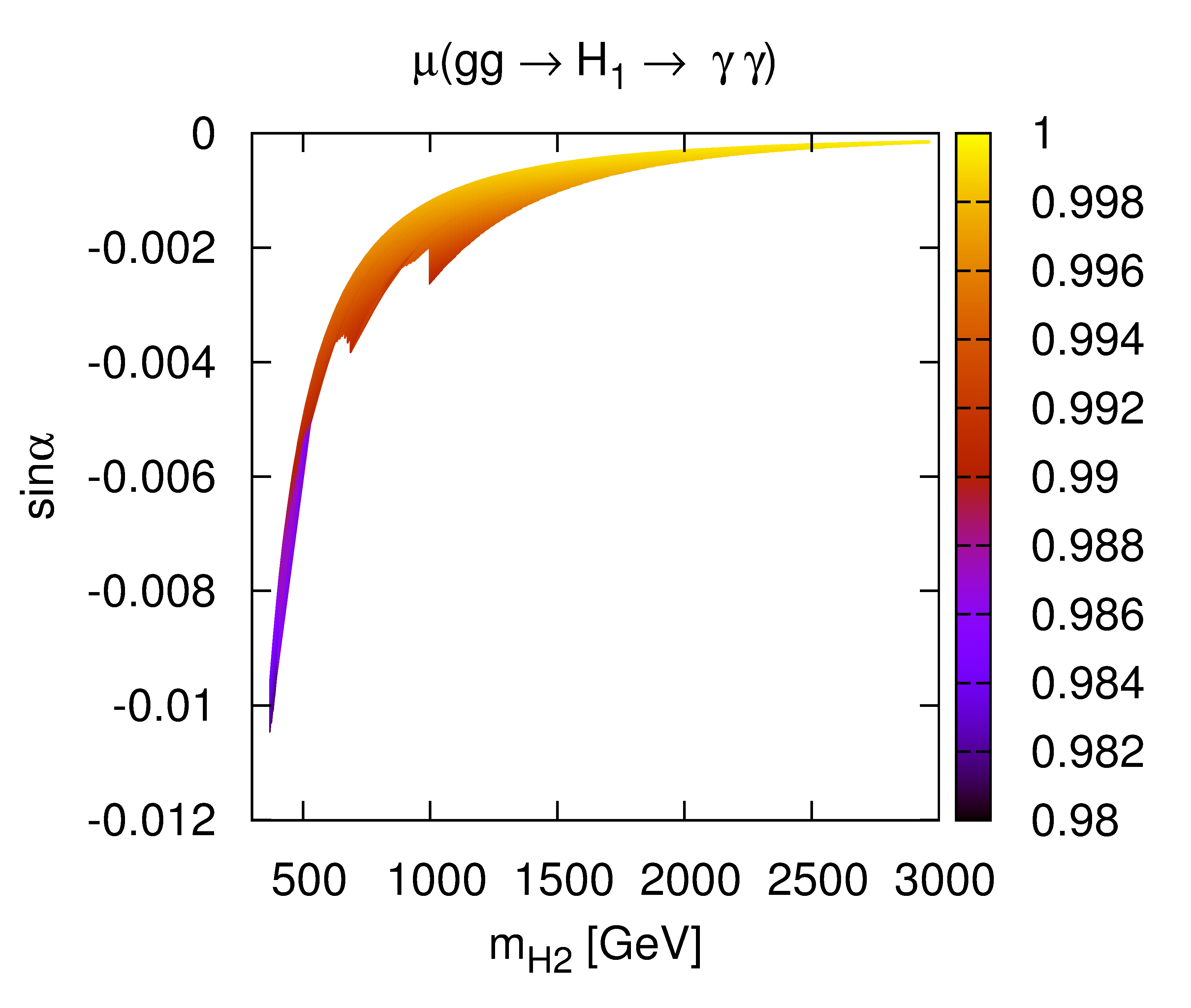}
\includegraphics[height=8cm,width=8cm]{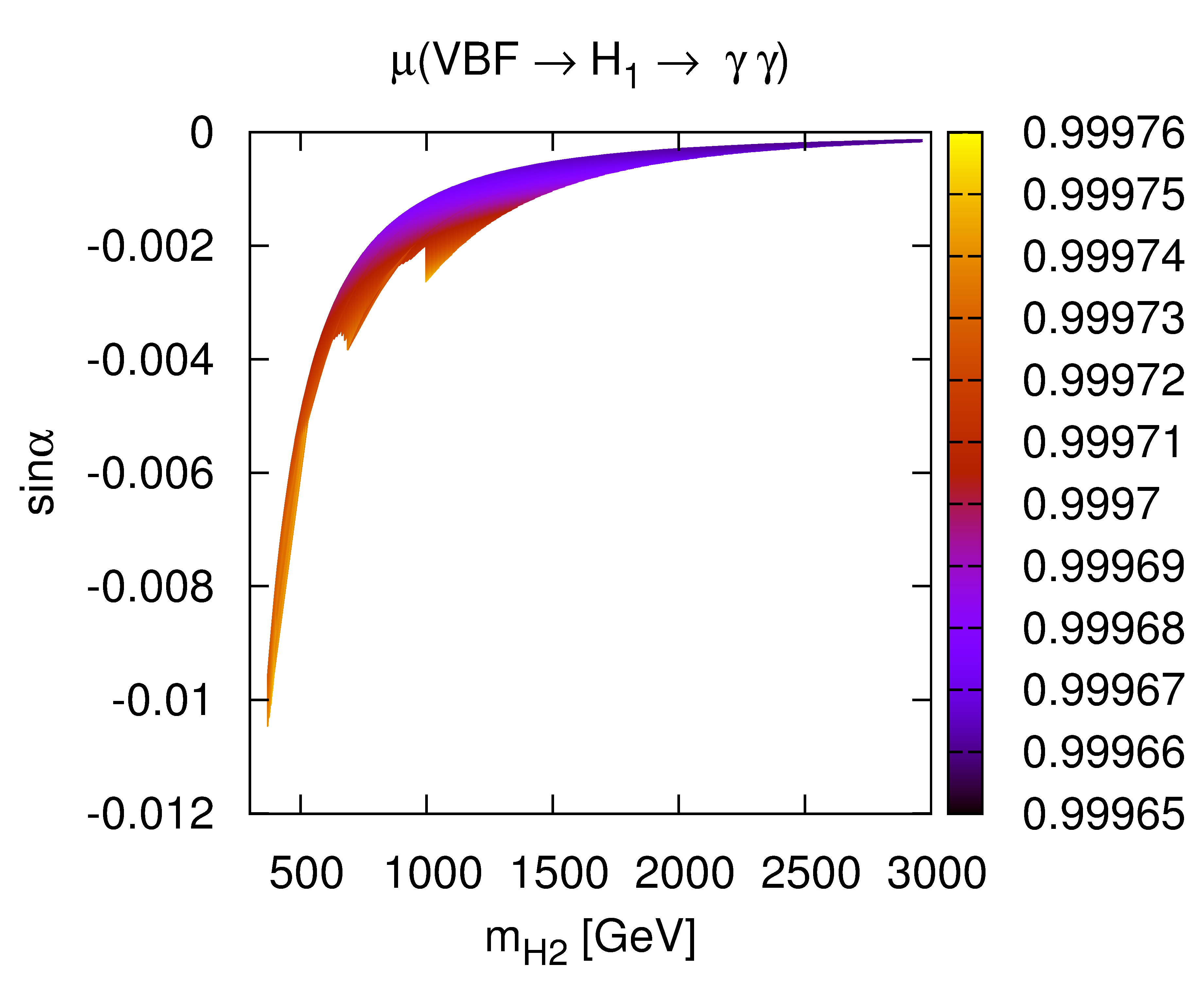}
\caption{Contour plots for the signal strengths for diphoton productions:
 ggF (left) and VBF (right) initiated processes. $H_1$ is fixed as $m_{H_1}=126$ GeV boson.}
\label{fig:AA}
\end{figure*}

We select the allowed parameter region by $3\sigma$ range
from the $\chi^2$ minima for CMS and ALTAS results.  In this  case there are another experimental exclusion bounds on the Higgs-like heavy mode by CMS and ATLAS 
with range up to $\sim 1000$ GeV \cite{CMS-HIG-12-034,Aad:2012uub,ATLAS13-067}.
In this case, the allowed range for the mixing angle $\alpha$ is severely restricted 
around the SM values, $\alpha \simeq 0$ (see Fig.\ref{fig:AA}). 

For both cases the signal strengths are very close to 1, the SM values. 
This means that the experimental data strongly favor the SM case and the dilaton 
should be heavy enough to decouple from the theory. 
Compared to the SM, only the surviving region for the heavier scalar mass $m_{H_2}$ 
is relatively relaxed compared with the constraints on the SM-like  Higgs boson, 
which can be expected because of the mixing between the SM Higgs boson $h$ and 
the dilation $\phi$ depending on the $( m_\phi , f_\phi )$ parameter values.
As a result, other observables as triple and quartic couplings are also strongly restricted 
around the SM values. 

As a result, unlike the Case I, it is not sufficient just to look into the observed 
$H_1 (m_{H_1}=126~{\rm GeV})$ sector to discriminate the model from the SM, 
since the model is pointing to the almost exact SM values for it.  The heavier scalar 
boson mass is constrained to be larger than $\sim$ 367 GeV from the Higgs signal 
strengths of the observed 126 GeV boson and the heavier Higgs searches (see Fig.~5).
This is a distinctive feature of our model compared with the SM. 
So the more detailed studies on the possible extra heavy scalar boson are necessary 
in the future $14$ TeV LHC and tentative International Linear Collider (ILC) to test 
this model more completely.

\section{Conclusions}

In this letter, we considered the SM coupled with some spontaneously broken scale 
symmetric sector with light dilaton (pseudo Nambu-Goldstone boson) 
(or the radion in the RS scenario) using the $SU(3)_C \times SU(2)_L \times U(1)_Y$ 
invariant form of the trace of  the energy momentum tensor of the SM fields, Eq.~(3).   
Our approach is different from others in that most earlier studies 
used the $T_\mu^\mu$ which is invariant under  $SU(3)_C \times U(1)_{\rm em}$ invariant form [ Eq.~(1) ], not the form which is invariant under the full SM gauge symmetry.  

The SM Higgs boson and the dilaton $\phi$  mix with each other after EWSB.
Since the original dilaton is coupled to the SM fields only through the Higgs mass 
parameter $\mu_H^2$ term and the quantum scale anomalies, two scalar bosons 
after the mixing carry the nature of the original dilaton and the SM Higgs boson.

Considering the $3\sigma$ ranges around the $\chi^2$ minima and experimental constraints on the extra 
light/heavy mode by LEP/LHC, the allowed region on the mixing angle and extra scalar mass is highly restricted.
 For the case of $126$ GeV boson and extra light scalar particle, 
we can give robust prediction for the mass of the extra light scalar and mixing angle. 
Also the correlations between the various signal strengths could be the 
good distinctive signals of our model. The triple and quartic couplings and their correlation give the 
impressive testbed for the model, which can be further studied in the 14 TeV LHC and ILC.

On the other hand, if we identify the observed scalar particle with mass $126$ GeV as 
light mode $H_1$, with the constraints upon the extra heavy SM-like scalar mode 
searched by CMS and ATLAS, the remaining parameter sets become severely confined 
around the SM expectations. This means that it is not enough to discriminate 
the model from the SM just by looking into the $126$ GeV sector. In this case, 
the more detailed study on the extra heavy mode will be necessary to test the model 
completely. 


\appendix
\section{The $\beta$ functions for the SM gauge group }
The $\beta$-functions that contribute to scale anomalies are collected for convenience~\cite{Arason:1991ic}:
\begin{eqnarray}
\frac{d g_i}{d t} &=& \beta_{g_i} = -b_i \frac{g_i^3}{16\pi^2}, \nonumber \\
\frac{d \bf{Y_u}}{d t} &=& \beta_{u} = \frac{\bf{Y_u}}{16\pi^2} 
  \left( \frac{3}{2}\left(\bf{Y_u}^\dagger\bf{Y_u}-\bf{Y_d}^\dagger\bf{Y_d} \right)
         +Y_2(S) -\left(\frac{17}{20}g_1^2+\frac{9}{4}g_2^2+8g_3^2 \right)
  \right), \nonumber \\
\frac{d \bf{Y_d}}{d t} &=& \beta_{d} = \frac{\bf{Y_d}}{16\pi^2} 
  \left( \frac{3}{2}\left(\bf{Y_d}^\dagger\bf{Y_d}-\bf{Y_u}^\dagger\bf{Y_u} \right)
         +Y_2(S) -\left(\frac{1}{4}g_1^2+\frac{9}{4}g_2^2+8g_3^2 \right)
  \right), \nonumber \\
\frac{d \bf{Y_e}}{d t} &=& \beta_{l} = \frac{\bf{Y_e}}{16\pi^2} 
  \left( \frac{3}{2} {\bf{Y_e}}^\dagger{\bf{Y_e}}
        -\frac{9}{4}\left(g_1^2+g_2^2\right)
  \right),
\end{eqnarray}
where
\begin{eqnarray}
b_Y&=&\left(-\frac{2}{3} n_f-\frac{1}{10}\right) \frac{5}{3}, \nonumber \\
b_2&=&-\frac{2}{3} n_f+\frac{22}{3}-\frac{1}{6}, \nonumber \\
b_3&=&-\frac{2}{3} n_f+11,\nonumber \\
Y_2(S)&=&\rm{Tr}\{3\bf{Y_u}^\dagger\bf{Y_u}+3\bf{Y_d}^\dagger\bf{Y_d}+\bf{Y_e}^\dagger\bf{Y_e}\}.
\end{eqnarray}

\begin{acknowledgments}
We thank Sunghoon Jung for useful discussion.
This work is supported in part by the National Research Foundation of Korea (NRF) 
Research Grant 2012R1A2A1A01006053 (P.K. and D.W.J.), and by the NRF grant funded 
by the Korea government (MSIP) (No. 2009-0083526) through  Korea Neutrino Research 
Center at Seoul National University (P.K.).
\end{acknowledgments}


\end{document}